# Temperature dependence of spin-transfer-induced switching of nanomagnets

I. N. Krivorotov, N. C. Emley, A. G. F. Garcia, J. C. Sankey, S. I. Kiselev, D. C. Ralph and R. A. Buhrman
*Cornell University, Ithaca, NY 14853*
(Dated March 30, 2004)

We measure the temperature, magnetic-field, and current dependence for the switching of nanomagnets by a spin-polarized current. Depending on current bias, switching can occur between either two static magnetic states or a static state and a current-driven precessional mode. In both cases, the switching is thermally activated and governed by the sample temperature, not a higher effective magnetic temperature. The activation barriers for switching between static states depend linearly on current, with a weaker dependence for dynamic to static switching.

The interaction between the magnetic moment of a metallic ferromagnet and a spin-polarized electrical current results in the spin-transfer effect [1-8], whereby the current can apply a torque to the magnet via transfer of angular momentum. The manipulation of nanomagnets by spin-transfer torques is under investigation for use in the switching of nonvolatile memory elements [9] and for current-tunable microwave sources [10,11]. Previous measurements of magnetic switching driven by spin-polarized currents have suggested that the process is thermally activated [12-14], but there remains disagreement about the switching mechanism. One set of models describes the effect of a spin-polarized current in terms of a torque that rotates the local moment of the magnet uniformly, as described within the framework of the Landau-Lifshitz-Gilbert (LLG) equation [1,15-17]. An alternative model proposes that when the polarization of the current is opposite to the moment of the magnet, spin-flip scattering of electrons excites non-uniform magnons, effectively raising the magnetic temperature so as to accelerate switching [13,14]. In this Letter, we use measurements of the switching rates as a function of temperature, magnetic field, and current to distinguish between these mechanisms. We find that a single sample can undergo different switching processes between separate static and dynamic states, that were not all distinguished in previous studies. In all cases, switching is thermally activated and governed by the actual background sample temperature. We observe no magnetic-configuration-dependent heating. The data are described well by current-dependent activation barriers that agree with predictions of the LLG-based models.

The samples we study are made from magnetic multilayers deposited by magnetron sputtering, with the structure Cu(100 nm)/ Py($X$ nm)/ Cu(6 nm)/ Py(2 nm)/ Cu($Y$ nm)/ Pt(30 nm), where $X$ = 12 or 20 nm for the thicker permalloy (Py = $Ni_{80}Fe_{20}$) layer and $Y$ = 2 or 20 nm. Electron-beam lithography and ion milling are used to define an elliptical pillar structure with a size 130 nm × 60 nm and with both magnetic layers etched through [5,18]. Top contact is made with a deposited Cu electrode. We will analyze the switching properties of the 2-nm thick Py "free layer". Positive currents are defined so that electrons flow from the thinner to thicker

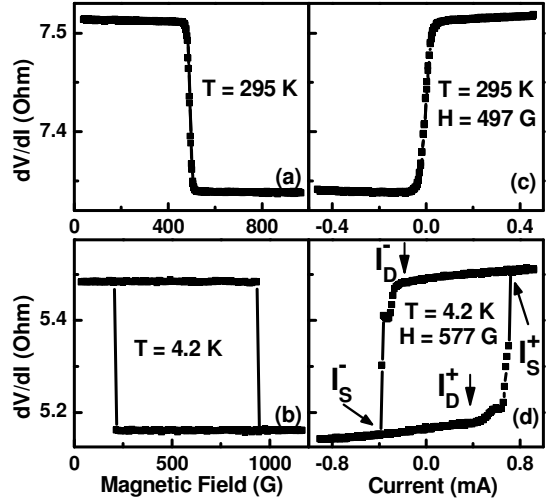

Fig. 1. (a,b) Differential resistance of a nanopillar spin valve device as a function of magnetic field and (c,d) as a function of current, measured at (a,c) $T$ = 295 K and (b,d) $T$ = 4.2 K.

Py layer. Although we focus below on one Py(20 nm)/Cu(6 nm)/Py(2 nm) device, similar results were obtained in eight samples, with three measured in detail.

Figure 1(a) displays the differential resistance ($dV/dI$) versus magnetic field ($H$) applied in plane along the major axis of the nanopillar, at the temperature $T$ = 295 K. The field drives the free layer moment between a low-resistance orientation parallel (P) to the fixed Py layer and a higher-resistance antiparallel (AP) state. The transition is continuous and reversible at 295 K and hysteretic at 4.2 K (Fig. 1(b)), indicating that the free layer is superparamagnetic at 295 K. The dipolar field from the fixed Py layer ($H_d$) causes the midpoint of the hysteresis loop to be shifted from $H$=0. Figures 1(c) and 1(d) show that the orientation of the free-layer moment can be controlled by the current $(I)$ as well as $H$. At 295 K, only 30 µA is required to saturate the free layer into either the P or AP state and thus suppress superparamagnetism. At 4.2 K the current hysteresis loop is not square, unlike the field loop. Starting in the low-resistance P state, as $I$ is increased there is a continuous increase in $dV/dI$ at $I_D^+$ prior to the abrupt switching to the AP state at $I_S^+$ [10,13]. From microwave measurements [10] on similar samples, we have identified this increase as due to the excitation of dynamical states (D) in which the free layer undergoes

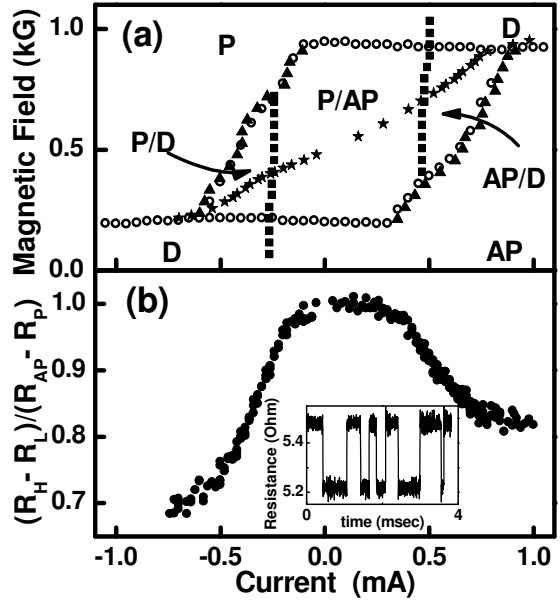

Fig. 2.(a) Phase diagram of magnetic states for the Py free layer at $T = 4.2$ K obtained from measurements of $dV/dI$ as a function of $I$ at fixed $H$ (■, $I_D^\pm$; ▲, $I_S^\pm$) and as a function of $H$ at fixed $I$ (○). Stars (★) mark $I$ and $H$ values for which the dwell times in both the high and the low resistance states are approximately equal to 1 ms, at sixteen temperatures ranging from 300 K at small $|I|$ to 4.2 K at large $|I|$. (b) The current dependence of the amplitude of two-level resistance fluctuations, normalized by the full difference in resistance between parallel and antiparallel states of the nanopillar spin valve. Inset: Two-level fluctuations between the high and low resistance states for $I = 1.0$ mA, $H = 956.6$ G, $T = 4.2$ K.

steady-state precessional motion.

Based on 4.2 K measurements of $dV/dI$ as a function of $I$ at a fixed $H$ and as a function of $H$ at a fixed $I$, we construct the phase diagram shown in Fig. 2(a), indicating at what values of $H$ and $I$ the different static and dynamic states are stable or bistable. This phase diagram is in good agreement with numerical [10] and analytical [19] solutions of the LLG equations. The phase diagram is shifted along the $H$ axis due to $H_d$, which allows us to access the regime where the total magnetic field ($H-H_d$) acting on the free layer is opposite to the fixed layer moment and thus to the current polarization.

Throughout the temperature range between 4.2 K and 300 K, $I$ and $H$ can be adjusted to bias points at which the sample resistance exhibits telegraph-type two-level fluctuations as a function of time, corresponding to transitions of the Py free layer between high (HR) and low (LR) resistance states (e.g., Fig. 2(b), inset). The bias conditions for which the dwell times in both HR and LR states are approximately equal to 1 ms are plotted in Fig. 2(a) for sixteen values of $T$. In Fig. 2(b), we show the difference in DC resistance between the HR ($R_H$) and the LR ($R_L$) states of the telegraph signal, normalized by the full resistance difference ($R_{AP}-R_P$) between the AP and P states for $I=0$ at the measurement temperature. We find that switching between the fully P and AP states occurs only for currents between -0.2 mA and 0.4 mA (Fig. 2(b)). The smaller changes in $R$ elsewhere indicate that for $I < -0.2$ mA the telegraph signals correspond to transitions between the P state and the dynamical state (D) with intermediate resistance, and for $I > 0.4$ mA the transitions are between the AP and D states. These identifications are consistent with the positions of the bistable modes in the 4.2 K phase diagram (Fig. 2(a)) and with the predicted phase diagram [19].

Before we discuss how the measured switching rates depend on $I$, $H$, and $T$, we will review the competing predictions. In general, the dwell time of a thermally-activated switching process can be parameterized in the form

$$\tau = \tau_0 \exp\left(\frac{E_a(H,I)}{k_B T^*}\right), \qquad (1)$$

where $\tau_0$ is an attempt time, $T^*$ is an effective temperature, $E_a(H,I)$ is an effective activation barrier, and $k_B$ is the Boltzmann constant. Within the model of coherent spin-transfer torques, analyzed in the framework of the LLG equation [12,17,20,22], $T^*$ is simply the true sample temperature, $T$. Even though the spin-transfer torques are non-conservative, an argument based on the fluctuation-dissipation theorem predicts that the effect of $I$ in this model for a uniaxial magnet undergoing P/AP switching can nevertheless be understood as a linear change in the effective activation barrier, $E_a(I) \propto 1-I/I_C$ [20,21]. The same functional form has also been found for the more general case both analytically and from numerical solutions of the LLG equation [17,22]. In contrast, quite different predictions are given by the model in which a spin-polarized current excites incoherent short-wavelength magnons and affects magnetic reversal by raising the effective temperature $T^*(I)$ of the free layer. Since spin-flip scattering is only enhanced when the current polarization is opposite to the moment of the free layer, this model predicts an increased $T^*(I)$ only for the P state (for positive $I$) or the AP state (for negative $I$), but not for both states simultaneously. The degree of heating has been argued to be very large and to increase with increasing $|I|$, e.g. 400 K/mA above a threshold current for devices in ref. [13], and 500 K to 1100 K in ref. [14]. In the magnetic heating model, $E_a(H,I)$ should depend only weakly on the $I$ through a decrease of the magnet's moment with increasing $T^*(I)$.

In Fig. 3 we show measured average dwell times, $\tau$, for the HR and LR states of the telegraph signals for negative (a,c) and positive (b,d) currents, plotted logarithmically as a function of $1/T$. Figures 3(a) and (b) correspond to high $T$ (200 K - 300 K), in the range of currents where transitions are between the static P and AP states, while Figs. 3(c) and (d) correspond to lower $T$ (4.2 K - 25 K) for switching between P and D states and between D and AP states. For each set of data corresponding to a given value of $I$, we first tuned $H$ so that $\tau$ in the two states were approximately equal, and then varied $T$ at fixed $I$ and $H$. Several conclusions can be drawn from these data. First, for all transitions at $T \geq$ 20 K, $\ln(\tau)$ depends approximately linearly on $1/T$, so that the transitions are thermally activated. As $H$ and $I$ are varied, the slopes of the lines change, meaning that the transitions remain thermally activated but the effective activation barriers are modified. At low $T$ and large $I$, the D→P and P→D dwell times (Fig. 3(c)) display an almost identical dependence on $T$, remaining approximately equal even as they both vary by several orders of magnitude. Similar behavior is also seen for D→AP and AP→D transitions (Fig. 3(d)) and the P→AP and AP→P transitions (Fig. 3(a)). This shows, in contrast to the magnetic heating model, that the free layer is not heated to a high effective temperature that depends on the orientation of its moment relative to the direction of the current polarization. The different slopes for P→AP and AP→P transitions seen in Fig. 3(b) are not due to heating, but to different sensitivities to $T$-dependent magnetic parameters (see Eq. (3) and discussion below).

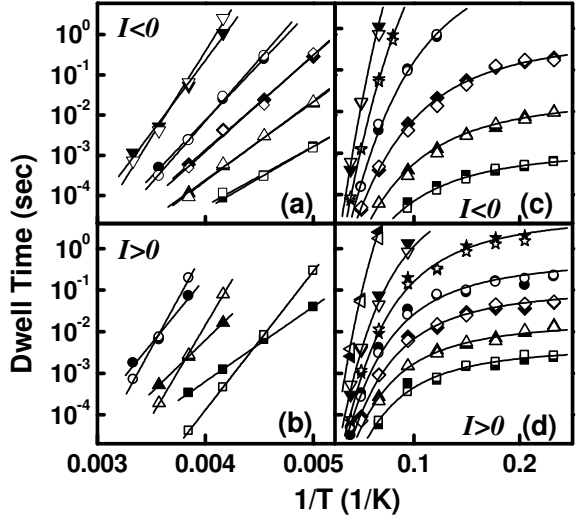

Fig. 3. Temperature dependence of dwell times for the two-level resistance fluctuations measured at fixed $I$ and $H$. (a) AP→P (open symbols) and P→AP (solid symbols) dwell times: (□,■) -0.26 mA, 400.8 G; (△,▲) -0.22 mA, 415.3 G; (◇,◆) -0.18 mA, 430.7 G; (○,●) -0.14 mA, 446.4 G; (▽,▼) -0.04 mA, 482.8 G. (b) AP→P (open symbols) and P→AP (solid symbols) dwell times: (□,■) 0.46 mA, 709.2 G; (△,▲) 0.36 mA, 649.4 G; (○,●) 0.16 mA, 561.6 G. (c) P→D (open symbols) and D→P transitions (solid symbols) dwell times: (□,■) -0.72 mA, 213.1 G; (△,▲) -0.70 mA, 217.8 G; (◇,◆) -0.68 mA, 222.6 G; (○,●) -0.66 mA, 226.8 G; (☆,★) -0.64 mA, 230.9 G; (▽,▼) -0.62 mA, 235.7 G. (d) D→AP (open symbols) and AP→D (solid symbols) dwell times: (□,■) 0.98 mA, 953.5 G; (△,▲) 0.96 mA, 950.7 G; (◇,◆) 0.94 mA, 947.5 G; (○,●) 0.92 mA, 944.7 G; (☆,★) 0.90 mA, 941.3 G; (▽,▼) 0.88 mA, 936.9 G; (◁,◀) 0.84 mA, 928.2 G. The lines are fits using Eqs. (1), (2) and (3).

While the dwell times for transitions between static and dynamical states in Figs. 3(c) and (d) exhibit thermal-activation behavior linear in $1/T$ for $T \geq 20$ K, they saturate below ~ 10 K. We conclude that heating by the current becomes non-negligible in this regime. However, this degree of heating is expected simply from ohmic dissipation in a diffusive metal wire, without more complicated considerations involving magnetic excitations. If heat flow from the device is dominated by electronic conduction to the contacts rather than by phonon mechanisms, the maximum electronic temperature $T_{el}$ in a metal wire is

$$T_{el} = \sqrt{T^2 + \frac{3}{4}\left(\frac{e \cdot I \cdot R}{\pi \cdot k_B}\right)^2}, \quad (2)$$

where $e$ is the electron charge [23]. The lines in Figs. 3(c) and 3(d) are fits to the data using Eq. (1), with $T^* = T_{el}$ and with $E_a(H,I)$ and $\tau_0$ as fitting parameters. The quality of the fits supports the picture of simple ohmic heating. For our highest currents at 4.2 K, the largest $T_{el}$ that we measure is less than 20 K.

From linear fits to the high $T$ data shown in Fig. 3(a) and 3(b), one may attempt to determine $E_a(H,I)$ and $\tau_0$. However, this process can yield unphysical results (e.g. $\tau_0 < 10^{-17}$ s), because the slopes of the $\ln(\tau)$ as a function of $1/T$, and hence the activation barrier, are affected by the $T$ dependence of the sample's magnetic parameters, particularly in the high $T$ range. In order to obtain quantitative results for the activation barriers, we have therefore analyzed the dwell times in a way that takes into account these $T$-dependences. We find that the evolution of the activation barriers is consistent with the assumption that their dependence on $H$ and $I$ factors [12,20,22], in the form

$$E_S(H,I) = \varepsilon_S(I) \cdot \frac{H_K(T) \cdot m(T)}{2} \cdot \left(1 \pm \frac{H - H_d(T)}{H_K(T)}\right)^{3/2}, \quad (3)$$

where S = HR or LR labels the activation barrier for transitions out of the HR or LR states, $H_K(T)$ is the shape anisotropy field of the free layer, and $m(T)$ is the magnetic moment of the free layer. The plus sign corresponds to the LR→HR transition and the minus sign to the HR→LR transition. The functions $\varepsilon_S(I)$ characterize the current dependence of the activation energies; they satisfy a normalization condition $\varepsilon_S(I)=1$ so that we recover the Néel-Brown activation barrier for $I=0$ [24]. To determine the required $T$ dependences, we assume that both $m(T)$ and $H_K(T)$ are proportional to magnetization ($M$) of the free Py layer. We use SQUID magnetometry to measure $M(T)$ for a 2 nm Py film in a Cu/Py/Cu trilayer that is exposed to the same heat treatments used during fabrication of the nanopillars [25]. The value of $H_K = 375$ G at 4.2 K is determined from the half-width of the hysteresis loop (Fig 1(b)) and the value of $m$ at 4.2 K is determined from $\varepsilon_S(I)=1$. We can measure $H_d$ directly for our device from variations of the hysteresis loop shift with $T$ (e.g., Fig. 1(a), 1(b)). We find a 19% decrease of $M(T)$ from 20 K to 300 K

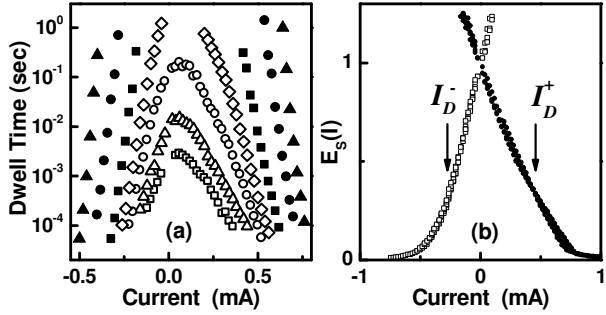

Fig. 4. (a) Measured dwell times as a function of $I$ at the field $H_{eq}(I,T)$ for which the average dwell times for LR→HR and HR→LR transitions are equal, for the temperatures □ 300 K, △ 280 K, ○ 260 K, ◇ 240 K, ■ 200 K, ● 140 K, and ▲ 80 K. (b) The activation barriers □ $\varepsilon_{HR}(I)$ for switching HR→LR and ● $\varepsilon_{LR}(I)$ for switching LR→HR, obtained by using Eq. (4) to collapse data such as in (a) for sixteen temperatures between 4.2 and 300 K.

[25] while $H_d$ decreases by 14% from 4.2 K to 295 K. The lines in Fig. 3(a) and 3(b) are fits to the data using Eqs. (1)-(3) with $\varepsilon_S(I)$ and $\tau_0$ as fitting parameters, yielding physically reasonable attempt times ($\tau_0 = 10^{-9.0 \pm 1.5}$ s). The different slopes for the two dwell times in Fig. 3(a) and 3(b) originate from the $T$-dependence of $|H-H_d|/H_K$ as described by Eq. (3). This parameter is a stronger function of $T$ for $H-H_d > 0$, since $|H-H_d|$ increases and $H_K$ decreases with $T$ (Fig. 3(b)), and a weaker function of $T$ for $H-H_d < 0$, since both $|H-H_d|$ and $H_K$ decrease with $T$ (Fig. 3(a)).

By combining Eqs. (1)-(3), and assuming that $\tau_0$ in the HR and LR states are approximately equal, we arrive at a simple expression for $\varepsilon_S(I)$:

$$\varepsilon_S(I) = \frac{2k_B T_{el}[\ln(\tau_S) - \ln(\tau_0)]}{H_K(T) \cdot m(T) \cdot \left(1 \pm \frac{H - H_d(T)}{H_K(T)}\right)^{3/2}}. \quad (4)$$

We can use Eq. (4) to rescale the dwell times measured at different values of $H$, $I$, and $T$ (Fig. 4(a)), onto common curves for $\varepsilon_{HR}(I)$ and $\varepsilon_{LR}(I)$ (Fig. 4(b)). The only unknown parameter in the analysis is $\tau_0$. A value of $\tau_0 = 10^{-9}$ s gives the best collapse of data sets at sixteen different temperatures onto just two curves for $\varepsilon_{HR}(I)$ and $\varepsilon_{LR}(I)$. The high quality of the data collapse in Fig. 4(b) provides justification for the form of the current-dependence asserted in Eq. (3), and it shows that the effect of a spin-polarized current on magnetic switching can be described accurately in terms of current-dependent activation barriers for the magnetic states. In the range of $I$ where transitions occur between static P and AP states, the activation barriers depend linearly on $I$ to good accuracy, as predicted by models of coherent spin-transfer torques that use the LLG equation [20,22]. However, at $|I| > I_D$, for transitions from the dynamical modes to a static state, the activation barriers appear to be weaker functions of $I$.

The value of $m$ at 4.2 K determined by our analysis is $m = (5.1 \pm 0.8) \times 10^{-15}$ emu. This can be compared to the expected value $m = M \cdot V = 7.9 \times 10^{-15}$ emu, where $M = 645$ emu/cm$^3$ at 20 K obtained from SQUID magnetometry of a test film [25] and $V = 1.23 \times 10^{-17}$ cm$^3$ is the estimated volume of the nanomagnet based on electron microscopy. We conclude that the activation volume of the free layer for thermally assisted reversal is close to its physical volume and thus the nanomagnet behaves as a single-domain Néel-Brown nanoparticle [26] from 4.2 K to 300 K.

In conclusion, we have performed measurements of magnetic switching rates for a nanomagnet under the influence of a spin-polarized current. The data are in good agreement with the spin-torque model, in which spin transfer causes a uniform rotation of the local magnetic moment that can be modeled by the LLG equation. The data are not consistent with arguments that incoherent magnon generation can drive a nanomagnet to effective temperatures well above room temperature. The overall effect of the current on magnetic switching is well-described in terms of current-dependent activation barriers.

We acknowledge support from DARPA through Motorola, from the Army Research Office, and from the NSF/NSEC program through the Cornell Center for Nanoscale Systems. We also acknowledge NSF support through use of the Cornell Nanofabrication Facility/NNIN and the Cornell Center for Materials Research facilities.